\documentclass{iopconfser}

\usepackage{amsmath}
\renewcommand{\vec}[1]{\mathbf{#1}}
\usepackage{braket}

\begin{document}

\title{Beyond Born-Oppenheimer Green's function theories: absolute and relational}

\author{Ville J Härkönen}

\affil{Computational Physics Laboratory, Tampere University, P.O. Box 692, FI-33014, Tampere, Finland}

\email{ville.j.harkonen@gmail.com}

\begin{abstract}
We consider quantum field theoretic many-body Green's function approach to solve the Coulomb many-body problem. The earlier beyond Born-Oppenheimer Green's function theories are absolute in nature and are based on the non-reduced Hamiltonian. Motivated by the issues following this approach we have developed a reduced Green's function theory which is relational in nature. The central differences between these approaches trace back to the relational-absolute debate, which has continued for over two thousand years since the time of Aristotle and still persists today. We highlight that these aspects of the theories are connected to several areas of modern physics, including relational quantum mechanics, quantum reference frames, superselection rules, the separation of different types of motion and spontaneous symmetry breaking. The starting point of any such exact theory must be absolute, given that the global conservation laws hold. However, for the observables to be meaningful, they must be defined with respect to the relative space. Approximations, such as the one introduced by Born and Oppenheimer, can break global symmetries and make certain parts of the absolute description well-defined. We highlight that spontaneous symmetry breaking is not necessary to explain the existence of solids, but entities such as phonons can be naturally explained within the relational theory.
\end{abstract}

\section{Introduction}
\label{Introduction}

The $n$-body problem of gravitation was formulated by Newton \cite{Newton-PhilosophiaNaturalisPrincipiaMathematica-1687}. This problem of classical mechanics has been a famously difficult problem to solve, and between 1750 and 1917, reportedly around 800 works focused on the three-body problem were published \cite{Whittaker-AtreatiseOnTheAnalyticalDynamicsOfParticlesAndRigidBodies-1917}, many of them by some of the most well-known scientists of their time, including Lagrange, Euler, and Poincar\'{e}. These efforts have also refined the approaches used in such a way that actually some of these results obtained are still of central importance in solving the quantum mechanical many-body problems. Interestingly, the many-body problem of electrons and nuclei interacting through Coulomb force resembles closely the $n$-body problem of gravitation and some part of the Hamiltonian function will be formally the same after changing the values of masses and constants in potential energies. Perhaps the most significant difference is that in the Coulomb problem attractive and repulsive interactions are present. This similarity of the many-body problem of electrons and nuclei and of the ancient $n$-body problem of gravitation leads to a similarity to theoretical tools and approaches used to solve these problems. Here we consider the quantum field theoretic many-body Green's function approach to solve the quantum mechanical many-body problem of electrons and nuclei. In the process, we will find that the results obtained for the gravitational $n$-body problem are very central to modern quantum mechanical theories and are linked to many facets of modern physics.

Consider the quantum mechanical problem of electrons and nuclei forming molecules or solids, with interactions originating from the experimentally verified static Coulomb force. In this case, the Hamiltonian operator can be written as
\begin{equation}
H = T_{e} + T_{n} + V_{ee} + V_{en} + V_{nn},
\label{eq:IntroductionEq_1}
\end{equation}
where $T_{e}$ is the kinetic energy for electrons, $T_{n}$ is the kinetic energy for nuclei, $V_{ee}$, $V_{en}$ and $V_{nn}$, the Coulomb potential energies for electron-electron, electron-nucleus and nucleus-nucleus interactions, respectively. The time-dependent Schrödinger equation separates into two components: one involving only the time variable and the other involving the electron and nuclear variables. The latter, time-independent Schrödinger equation, is
\begin{equation}
H \Psi\left(\vec{r},\vec{R}\right) = E \Psi\left(\vec{r},\vec{R}\right),
\label{eq:IntroductionEq_2}
\end{equation}
where $\vec{r} \equiv \vec{r}_{1}, \ldots, \vec{r}_{N_{e}}$ denotes all the $N_{e}$ electron coordinates and $\vec{R} \equiv \vec{R}_{1}, \ldots, \vec{R}_{N_{n}}$ all the $N_{n}$ nuclear coordinates. For the hydrogen atom, Eq. \ref{eq:IntroductionEq_2} with $N_{e} = N_{n} = 1$, can be solved with a pen and paper approach \cite{Dirac-PrinciplesOfQM-1958,Bethe-QuantumMechanicsOfOneAndTwoElectronAtoms-1957}. For larger $N_{e}$ and $N_{n}$, systems often considered are molecules and solids for which the general problem of Eq. \ref{eq:IntroductionEq_2} is difficult to solve in many of these cases, even numerically. The numerical difficulties in the case of molecules and solids are related to the bad scaling as a function of $N_{e},N_{n}$ rendering the wave function unsuitable for practical computations in many cases. The problem given by Eq. \ref{eq:IntroductionEq_2} was considered in the very early stages of quantum mechanics when Max Born and J. Robert Oppenheimer derived the famous Born-Oppenheimer (BO) approximation \cite{Born-OppenheimerAdiabaticApprox-1927,Born-Huang-DynamicalTheoryOfCrystalLattices-1954}. In the BO approximation, $\Psi\left(\vec{r},\vec{R}\right) \approx \Phi_{\vec{R}}\left(\vec{r}\right) \chi\left(\vec{R}\right)$, where the electronic $\Phi_{\vec{R}}\left(\vec{r}\right)$ and nuclear wave functions $\chi\left(\vec{R}\right)$ satisfy
\begin{align}
H_{BO} \Phi_{\vec{R}}\left(\vec{r}\right) &= \mathcal{E}_{BO}\left(\vec{R}\right) \Phi_{\vec{R}}\left(\vec{r}\right), \label{eq:IntroductionEq_3_1} \\
H_{n} \chi\left(\vec{R}\right) &= E \chi\left(\vec{R}\right). \label{eq:IntroductionEq_3_2}
\end{align}
Here, the electronic BO Hamiltonian is $H_{BO} \equiv H - T_{n}$ and the nuclear BO Hamiltonian is of the form $H_{n} \equiv T_{n} + \mathcal{E}_{BO}\left(\vec{R}\right)$. These two equations have been a central starting point in developing the understanding of the properties of molecules and solids for nearly 100 years. To be more specific, for molecules \cite{Born-OppenheimerAdiabaticApprox-1927} these equations appear a bit different and we will discuss this aspect more in Sec. \ref{SymmetryAndReduction}. The BO approximation is justified by the large mass difference of electrons and nuclei, and it is expected to be relatively accurate for a majority of molecules and solids. However, particularly the electronic problem, Eq. \ref{eq:IntroductionEq_3_1}, still suffers from the mentioned dimensionality problem, which creates a need for alternative approaches. Such approaches have been developed, including the density functional theory \cite{Hohenberg-DFT-PhysRev.136.B864-1964,KohnSham-DFT-PhysRev.140.A1133-1965,DreizlerGross-DFTbook-1990} and the method of many-body Green's functions \cite{Gross-ManyParticleTheory-1991}. The latter approach adopt the tools from quantum field theory and in the past few decades the BO Green's function approach has been systematically used to describe molecules and solids \cite{Golze-TheGWCompendiumAPracticalGuideToTheoreticalPhotoemissionSpectroscopy-2019}.

But all the aforementioned alternative approaches to describe observables of the many-body problem defined by the Hamiltonian of Eq. \ref{eq:IntroductionEq_1} are assuming the BO approximation. There are some systems where the BO approximation may fail \cite{Vidal-EvidenceOnTheBreakdownOfTheBornOppenheimerApproximationInTheChargeDensityOfCrystalline7LiHD-1992} and in these case we have to start from the beginning, from Eq. \ref{eq:IntroductionEq_1}. Yet again, one such approach is the many-body Green's function theory \cite{Baym-field-1961,vanLeeuwen-FirstPrincElectronPhonon-PhysRevB.69.115110-2004,Harkonen-ManyBodyGreensFunctionTheoryOfElectronsAndNucleiBeyondTheBornOppenheimerApproximation-PhysRevB.101.235153-2020}. In this work, we discuss some of the developments of beyond-BO Green's function theories over the past six decades or so. Several interesting and subtle issues remain with these approaches, and the first implementation of beyond-BO Green's function theory is yet to be done.

Before discussing the developed beyond-BO Green's function theories in Sec. \ref{GreensFunctionTheories}, we address an important aspect of these theories in Sec. \ref{SymmetryAndReduction}: namely, the symmetries of the Hamiltonian in Eq. \ref{eq:IntroductionEq_1} and their consequences. These seem to be connected to several facets of physics: reduction theory in classical and quantum mechanics, the absolute vs relational description of nature, the description of different types of motion (internal, rotational, translational), gauge theories, superselection rules, quantum reference frames and statements about what can actually be measured.

\section{Symmetry and Reduction}
\label{SymmetryAndReduction}

\subsection{Preface}
\label{Preface}

The so-called reduction is an important part of solving the $n$-body problem and as the name implies it involves reducing the number of degrees of freedom \cite{Whittaker-AtreatiseOnTheAnalyticalDynamicsOfParticlesAndRigidBodies-1917}. This is done by separating different types of motion in the system, such that, for instance, in the transformed coordinates, the motion is separated into the independent motion of the system's center of mass and its internal motion in systems where the external forces are absent. Since the center-of-mass motion is that of a single particle and independent from the internal motion, we can neglect these degrees of freedom and concentrate on the internal motion only, which, after the mentioned reduction, now has three degrees of freedom fewer than the original problem we started with. A further reduction can also be conducted and is related to the conservation of angular momentum. These developments in classical mechanics were perhaps the reasons why Born and Oppenheimer were using the reduction in solving the quantum mechanical many-body problem of Coulomb interaction. Namely, the center-of-mass motion was separated and the partial separation of rotational and internal motion was achieved by a subsequent transformation \cite{Born-OppenheimerAdiabaticApprox-1927}. Thus, the Hamiltonian of Eq. \ref{eq:IntroductionEq_1}, which we call from now on the non-reduced Hamiltonian, was not the actual Hamiltonian from which the observables in the original BO approach were computed, but rather the Hamiltonian reduced from Eq. \ref{eq:IntroductionEq_1} instead. The reduction theory of classical mechanics has been further developed and can be formulated in the language of modern differential geometry \cite{Marsden-IntroductionToMechanicsAndSymmetry-2013,Butterfield-OnSymplecticReductionInClassicalMechanics-2006}. As already in the original work by Born and Oppenheimer on molecules \cite{Born-OppenheimerAdiabaticApprox-1927}, the reduction theory is also used in the quantum mechanical context \cite{Hirschfelder-SeparationOfRotationalCoordinatesFromTheSchrodingerEquationForNParticles-1935,Wilson-MolecularVibrationsTheTheoryOfInfraredAndRamanVibrationalSpectra-1955,Sutcliffe-TheDecouplingOfElectronicAndNuclearMotions-2000}. Several interesting findings related to reduction have been made in the past few decades. For instance, the separation of rotational and internal motions has interesting connections to gauge theories, as the internal dynamics of a many-body system can be viewed as one \cite{Tachibana-CompleteMolecularHamiltonianBasedOnTheBornOppenheimerAdiabaticApproximation-PhysRevA.33.2262-1986,Littlejohn-GaugeFieldsInTheSeparOfRotatAndIntMotionsIntheNbodyProb-RevModPhys.69.213-1997,Littlejohn-GaugeTheoryOfSmallVibrationsInPolyatomicMolecules-2002}.

The reduction theory has a strong connection to the symmetries of the Hamiltonian, and this further seems to be connected to several aspects of quantum mechanics. The Hamiltonian in Eq. \ref{eq:IntroductionEq_1} is invariant under Galilean transformations. Specifically, it is invariant under translations and rotations of all particle coordinates, leading to the conservation of total momentum and total angular momentum of the system, respectively. There are several ways to state the interesting consequences of these symmetries. To summarize a central consequence, it appears difficult to extract meaningful physical information about the system if the non-reduced Hamiltonian in Eq. \ref{eq:IntroductionEq_1} is used as the basis of the theory as such. Nevertheless, the non-reduced Hamiltonian is a starting point of theories in the solid state \cite{Born-Huang-DynamicalTheoryOfCrystalLattices-1954}. As already mentioned, in molecules the theory derived by Born and Oppenheimer is based on the Hamiltonian reduced from Eq. \ref{eq:IntroductionEq_1} \cite{Born-OppenheimerAdiabaticApprox-1927}. The way we treat these symmetries and the reduction is also the main difference of the many-body Green's function theories. Namely, the first many-Body Green's function theory \cite{Baym-field-1961} and theories that have followed it \cite{Keating-DielScreeningAndThePhononSpectraOfMetalAndNonMetalCrystals-PhysRev.175.1171-1968,Gillis-SelfConsistentPhononsAndTheCoupledElectronPhononSystem-PhysRevB.1.1872-1970,Giustino-ElectronPhononInteractFromFirstPrinc-RevModPhys.89.015003-2017,Stefanucci-InAndOutOfEquilibriumAbInitioTheoryOfElectronsAndPhonons-PhysRevX.13.031026-2023}, are based on the non-reduced Hamiltonian. This leads to certain problems in the computation of observables and also in the description of different types of motion, which has been a motivation to develop Green's function theories based on the reduced Hamiltonian \cite{vanLeeuwen-FirstPrincElectronPhonon-PhysRevB.69.115110-2004,Harkonen-ManyBodyGreensFunctionTheoryOfElectronsAndNucleiBeyondTheBornOppenheimerApproximation-PhysRevB.101.235153-2020,Harkonen-ExactFactorizationOfTheManyBodyGreensFunctionTheoryOfElectronsAndNuclei-PhysRevB.106.205137-2022}. In this section we discuss the mentioned symmetry and reduction related topics from different points of view.

\subsection{Reduction}
\label{Reduction}

The issues related to symmetries are discussed extensively in the literature on molecules \cite{Sutcliffe-TheDecouplingOfElectronicAndNuclearMotions-2000}, as they are also related to separation of different types of motion through reduction theory, but are not hardly discussed or acknowledged at all in the solid state literature. Yet, further considerations are needed in the case of solids, even in the BO approximation, when meaningful observables are computed from Eq. \ref{eq:IntroductionEq_3_2}, for instance (see Sec. \ref{ConnectionsToDifferentFacetsOfPhysics}).

We first consider the translation symmetry of the non-reduced Hamiltonian of Eq. \ref{eq:IntroductionEq_1}. From the translational symmetry it follows that the non-reduced Hamiltonian commutes with the total momentum operator and we can choose the eigenstates of the Hamiltonian to be mutual eigenstates of the total momentum operator. However, such choice means that the eigenstates are not square integrable \cite{Sutcliffe-TheDecouplingOfElectronicAndNuclearMotions-2000}. All choices for the eigenbasis of the Hamiltonian should give the same expected values of observables, which creates an interesting consequence. Given the non-reduced Hamiltonian of Eq. \ref{eq:IntroductionEq_1} and its eigenstates from Eq. \ref{eq:IntroductionEq_2} we cannot compute meaningful observables. Another way to state this is that if we choose the eigenstates to be the mutual eigenstates of total momentum operator, we exactly know the momentum and we do not know anything about the position of the system in space. We thus obtain constant one-body particle densities for all species of particles. Quantum mechanics thus seems to indicate that we cannot extract meaningful physical observables from the eigenstates of the non-reduced Hamiltonian of Eq. \ref{eq:IntroductionEq_1}. This issue can be resolved by conducting reduction, the simplest system being the hydrogen atom. The original variables of the hydrogen atom are $\left(\vec{r}_{p},\vec{r}_{e}\right)$, where $\vec{r}_{p}$ is the position operator for proton and $\vec{r}_{e}$ for electron. The transformation can be written as $\left(\vec{r}_{p},\vec{r}_{e}\right) \rightarrow \left(\vec{r}_{cm},\vec{r}\right)$, where $\vec{r}_{cm}$ is the center-of-mass coordinate and $\vec{r} = \vec{r}_{e} - \vec{r}_{p}$ the relative coordinate of the electron and proton. The Hamiltonian transforms from $H = T_{e} + T_{p} + V$ to $H = T_{cm} + T_{r} + V$, where the center-of-mass kinetic energy $T_{cm}$ commutes with $T_{r} + V \equiv H'$ and we can separate the problem into two parts. The well known atomic orbitals of the hydrogen atom are solved from the Schrödinger equation of the relative coordinates with the Hamiltonian $H'$ \cite{Dirac-PrinciplesOfQM-1958,Bethe-QuantumMechanicsOfOneAndTwoElectronAtoms-1957}. An approximate solution can be obtained by assuming the BO approximation by taking the limit of infinite proton mass leading to vanishing proton kinetic energy. In this case, the Schrödinger equation for $H_{BO} = T_{e} + V$ can be solved for electron only for a fixed proton position.

The rotational symmetry leading to the conserved angular momentum requires a similar reduction procedure described in detail in the literature \cite{Hirschfelder-SeparationOfRotationalCoordinatesFromTheSchrodingerEquationForNParticles-1935,Wilson-MolecularVibrationsTheTheoryOfInfraredAndRamanVibrationalSpectra-1955,Sutcliffe-TheDecouplingOfElectronicAndNuclearMotions-2000,Littlejohn-GaugeFieldsInTheSeparOfRotatAndIntMotionsIntheNbodyProb-RevModPhys.69.213-1997,Littlejohn-GaugeTheoryOfSmallVibrationsInPolyatomicMolecules-2002}. To summarize, the reduction amounts a transformation $\left(\vec{r},\vec{R}\right) \rightarrow \left(\vec{r}',\vec{R}'\right)$, where the latter contain the internal coordinates (sometimes called the shape coordinates or relational degrees of freedom), three degrees of freedom describing the rotations and another three the center-of-mass coordinate of the whole system. As in the case of hydrogen atom, the center-of-mass coordinate can be completely separated from the other degrees of freedom while the rotations and internal coordinates in general cannot \cite{Littlejohn-GaugeFieldsInTheSeparOfRotatAndIntMotionsIntheNbodyProb-RevModPhys.69.213-1997}. The Hamiltonian can thus be written as \cite{Harkonen-ManyBodyGreensFunctionTheoryOfElectronsAndNucleiBeyondTheBornOppenheimerApproximation-PhysRevB.101.235153-2020}
\begin{equation}
H = T_{cm} + H',
\label{eq:ReductionEq_1}
\end{equation}
where $T_{cm}$ and $H'$ commute and $H'$ is the reduced Hamiltonian of the internal and (possibly) rotational degrees of freedom. The reduced Hamiltonian $H'$ does not have translational or rotational symmetries with respect to the new coordinates $\left(\vec{r}',\vec{R}'\right)$ and this Hamiltonian can be used to develop quantum mechanical theory of electrons and nuclei. Such a wave function approach was developed in the very early stages of quantum mechanics \cite{Born-OppenheimerAdiabaticApprox-1927} and have been developed further \cite{Littlejohn-GaugeFieldsInTheSeparOfRotatAndIntMotionsIntheNbodyProb-RevModPhys.69.213-1997,Sutcliffe-TheDecouplingOfElectronicAndNuclearMotions-2000}, but such a theory of many-body Green's functions for electrons and nuclei has been developed only recently \cite{Harkonen-ManyBodyGreensFunctionTheoryOfElectronsAndNucleiBeyondTheBornOppenheimerApproximation-PhysRevB.101.235153-2020}. Additionally, the so-called multicomponent density functional theory has also been developed \cite{Kreibich-MulticompDFTForElectronsAndNuclei-PhysRevLett.86.2984-2001} and is based on the reduced version of the theory with the Hamiltonian of Eq. \ref{eq:IntroductionEq_1} as a starting point.

\subsection{Connections to different facets of physics}
\label{ConnectionsToDifferentFacetsOfPhysics}

The necessity of reduction in obtaining reasonable physical quantities (when no additional approximations are made) in the quantum mechanical many-body problem appear to have many connections to various areas of physics. Here we discuss these connections and start with the spontaneous symmetry breaking. We can obtain the BO Hamiltonian $H_{BO}$ from Eq. \ref{eq:IntroductionEq_1} by setting $T_{n} = 0$ and thus Eq. \ref{eq:IntroductionEq_3_1} is a special case of Eq. \ref{eq:IntroductionEq_2}. The exact problem has the Galilean symmetry while the electronic BO equation can be seen to break these symmetries. Namely, the position variables $\vec{R}$ commute with the Hamiltonian $H_{BO}$ and can be treated as parameters. The BO wave function is normalized as $\int d\vec{r} \left|\Phi_{\vec{R}}\left(\vec{r}\right)\right|^2 = 1$ for given $\vec{R}$. In solids, for instance, the electron-nuclei potential included to $H_{BO}$ therefore brings in lattice periodicity for periodic arrangement of the nuclei $\vec{R}$ and Bloch theorem follows \cite{Kittel-QuantumTheoryOfSolids-1987}. Such periodic symmetry seems to be an artifact of the electronic BO approximation and without assuming this approximation the existence of periodic solids is sometimes explained through spontaneous symmetry breaking of continuous translational and rotational symmetries \cite{Beekman-AnIntroductionToSpontaneousSymmetryBreaking-2019}. We state that the spontaneous symmetry breaking is not necessarily needed for explaining the existence of solids and of phonons as Nambu-Golstone bosons. Namely, after the reduction, the Hamiltonian $H'$ does not have the translational or rotational symmetry anymore with respect to the variables $\left(\vec{r}',\vec{R}'\right)$ and for instance phonons or real normal modes can be derived from it and its eigenstates. In practical calculations on solids, periodic Born von-Karman boundary conditions are imposed \cite{Born-Huang-DynamicalTheoryOfCrystalLattices-1954}, which breaks the rotational symmetry. Imposing such a condition is an approximation, but in the case of solids sufficiently large, the error introduced can be justified to be relatively small \cite{Born-Huang-DynamicalTheoryOfCrystalLattices-1954} and the reduction related to the angular momentum conservation is not needed anymore. After the boundary condition, the system still has the continuous translational symmetry and indeed from this symmetry it follows that in the non-reduced formulation there will be three acoustic phonon modes of zero frequencies corresponding to the translation of the whole lattice. In the harmonic approximation, the nuclear wave function satisfying Eq. \ref{eq:IntroductionEq_3_2} can be written in normal coordinates $q$ as \cite{Harkonen-OnTheDiagonalizationOfQuadraticHamiltonians-2021} $\chi = \chi_{1}\left(q_{1}\right) \cdots \chi_{3N_{n}}\left(q_{3N_{n}}\right)$, where each single particle function is of a simple harmonic oscillator form and thus for instance for the ground state $\chi_{s}\left(q_{s}\right) = \left(\omega_{s}/\pi\right)^{1/4} \exp\left[-\omega_{s} q^{2}_{s}\right]$. Due to the translational symmetry three of the frequencies $\omega_{s}$ are zero meaning that $\chi = 0$ and thus also $\Psi\left(\vec{r},\vec{R}\right) = 0$. This example illustrates the issues present in the non-reduced theory. In some cases, these issues can be still avoided by simply neglecting the contributions from the zero frequency translational modes and quantities like mean square displacements of the nuclei can be successfully computed \cite{Harkonen-NTE-2014}. In general, a care must be taken to ensure that these contributions are correctly eliminated. In all cases, the reduction resolves these issues and three modes related to the center-of-mass degree of freedom can be eliminated.

The consequences of symmetry of the Hamiltonian of Eq. \ref{eq:IntroductionEq_1} can be also stated in terms of the so-called superselection rules \cite{Wick-TheIntrinsicParityOfElementaryParticles-PhysRev-1952,Aharonov-QuantumParadoxesQuantumTheoryForThePerplexed-2005,Bartlett-ReferenceFramesSuperselectionRulesAndQuantumInformation-RevModPhys.79.555-2007}. For example, the translational symmetry of the Hamiltonian leads to a superselection rule of momentum which has been claimed to prevent the observation of absolute position \cite{Aharonov-ChargeSuperselectionRule-PhysRev.155.1428-1967,Aharonov-ObservabilityOfTheSignChangeOfSpinorsUnder2piRotations-PhysRev.158.1237-1967,Page-EvolutionWithoutEvolutionDynamicsDescribedByStationaryObservables-PhysRevD.27.2885-1983}. Other way around, sometimes this is stated such that \cite{Bartlett-ReferenceFramesSuperselectionRulesAndQuantumInformation-RevModPhys.79.555-2007} a lack of (quantum) reference frame leads to a superselection rule which prevents the observation of absolute positions \cite{Page-EvolutionWithoutEvolutionDynamicsDescribedByStationaryObservables-PhysRevD.27.2885-1983}. Indeed, when we compute observables in the non-reduced system, lacking a frame of reference, we have seen that non-physical results will appear also in practice. The reduction on the other hand resolves these issues and can be seen as a way to make our quantum mechanical model relational \cite{Rovelli-RelationalQuantumMechanics-1996,Poulin-ToyModelForARelationalFormulationOfQuantumTheory-2006}. Our discussion on the Coulomb problem implies that relational variables are needed to obtain meaningful observables, if no further approximations are made. On the other hand, the BO approximation can be seen as to make the full quantum mechanical problem semi-classical \cite{Poulin-ToyModelForARelationalFormulationOfQuantumTheory-2006} and the electronic BO problem of Eq. \ref{eq:IntroductionEq_3_1} is free from the discussed issues.

\section{Green's Function Theories}
\label{GreensFunctionTheories}

\subsection{Born-Oppenheimer Theory}
\label{BornOppenheimerTheory}

To get a better overall view on the Green's function theories we summarize these approaches within the BO approximation \cite{Gross-ManyParticleTheory-1991}. Consider first the electronic BO problem of Eq. \ref{eq:IntroductionEq_3_1}. We first note that the Hamiltonian can be written as
\begin{equation} 
\hat{H}_{BO} = \hat{T}_{e} + \hat{V}_{ee} + \hat{V}_{en} + V_{nn},
\label{eq:BornOppenheimerTheoryEq_1}
\end{equation}
where (we have set $\hbar = 1$)
\begin{align}
\hat{T}_{e} &=  -\frac{ 1 }{ 2 m_{e} } \int d\vec{y} \hat{\psi}^{\dagger}\left(\vec{y}\right) \nabla^{2}_{\vec{y}} \hat{\psi}\left(\vec{y}\right), \nonumber \\
\hat{V}_{en} &= \int d\vec{y} \int d\vec{y}'  v\left(\vec{y}, \vec{y}'\right) n_{n}\left(\vec{y}'\right) \hat{\psi}^{\dagger}\left(\vec{y}\right) \hat{\psi}\left(\vec{y}\right), \nonumber \\
\hat{V}_{ee} &= \frac{1}{2} \int d\vec{y} \int d\vec{y}' v\left(\vec{y},\vec{y}'\right) \hat{\psi}^{\dagger}\left(\vec{y}\right) \hat{\psi}^{\dagger}\left(\vec{y}'\right) \hat{\psi}\left(\vec{y}'\right) \hat{\psi}\left(\vec{y}\right), \nonumber \\
V_{nn} &= \frac{1}{2} \sum^{N_{n}}_{ k,k'= 1 }{}^{'}  \frac{ Z_{k} Z_{k'} \varsigma }{ \left| \vec{R}_{k} - \vec{R}_{k'} \right| }.
\label{eq:BornOppenheimerTheoryEq_2}
\end{align}
Here, we denote $n_{n}\left(\vec{y}\right) = - \sum_{ k } Z_{k} \delta\left( \vec{y} - \vec{R}_{k} \right)$, $v\left(  \vec{y}, \vec{y}' \right) = \varsigma / \left| \vec{y} - \vec{y}' \right|$ and $\varsigma = e^{2}/ \left( 4 \pi \epsilon_{0} \right)$. The primed sum is over the nuclear labels $k \neq k'$ and the nuclear variables $\vec{R}$ can be treated as parameters in the electronic BO problem as they commute with $\hat{H}_{BO}$. The anti-commutation relations between the electronic field operators are
\begin{equation}
\left[\hat{\psi}\left(\vec{y}\right),\hat{\psi}^{\dagger}\left(\vec{y}'\right)\right]_{+} = \delta\left(\vec{y}-\vec{y}'\right).
\label{eq:BornOppenheimerTheoryEq_3}
\end{equation}
The electronic one-body BO Green's function is defined as
\begin{equation} 
G^{BO}_{\vec{R}}\left(\vec{y}t,\vec{y}'t'\right) \equiv \frac{1}{i} \frac{\text{Tr}\left[e^{-\beta \hat{H}^{M}_{BO}} \mathcal{T}\left\{ \hat{\psi}\left(\vec{y}t\right) \hat{\psi}^{\dagger}\left(\vec{y}'t'\right) \right\} \right]_{\Phi_{\vec{R}}}}{\text{Tr}\left[e^{-\beta \hat{H}^{M}_{BO}}\right]_{\Phi_{\vec{R}}}},
\label{eq:BornOppenheimerTheoryEq_4}
\end{equation}
where $\mathcal{T}\left\{\cdots\right\}$ denotes the time-ordering and $\hat{\psi}\left(\vec{y}t\right) \equiv \hat{U}^{\dagger}_{BO}\left(t\right) \hat{\psi}\left(\vec{y}\right) \hat{U}_{BO}\left(t\right)$ is an operator in the Heisenberg picture. The evolution operator $\hat{U}_{BO}\left(t\right)$ is defined with respect to the time-independent BO Hamiltonian $\hat{H}_{BO}$. We denoted $\beta = k^{-1}_{B} T^{-1}$. Further, $\hat{H}^{M}_{BO} \equiv \hat{H}_{BO} - \mu_{e} \hat{N}_{e}$, where $\mu_{e}$ is the chemical potential of the electrons and $\hat{N}_{e}$ the electron number operator. The trace in Eq. \ref{eq:BornOppenheimerTheoryEq_4} is of the form $\text{Tr}\left[\hat{o}\right]_{\Phi_{\vec{R}}} = \sum_{m} \braket{\Phi^{\left(m\right)}_{\vec{R}}|\hat{o}|\Phi^{\left(m\right)}_{\vec{R}}}$, where $m$ labels the electronic BO states and $\hat{o}$ is an operator acting in the electronic Hilbert space. Through the trace, the electronic BO Green's function in Eq. \ref{eq:BornOppenheimerTheoryEq_4}, is parametrically dependent on the nuclear variables $\vec{R}$.  We can solve $G^{BO}_{\vec{R}}\left(\vec{y}t,\vec{y}'t'\right)$ by using the many-body perturbation theory or the EOM approach \cite{Stefanucci-Leeuwen-many-body-book-2013}. In both of these approaches further approximations are needed. The theory of Green's function $G^{BO}_{\vec{R}}\left(\vec{y}t,\vec{y}'t'\right)$ is rather well-known and has become a valuable computational tool in the description of realistic materials \cite{Golze-TheGWCompendiumAPracticalGuideToTheoreticalPhotoemissionSpectroscopy-2019}.

The many-body Green's function theory for the nuclei in the BO approximation \cite{Maradudin-Fein-PhysRev.128.2589-Scat-Neutr.1962} is in a similar way based on the Hamiltonian of the nuclear equation of Eq. \ref{eq:IntroductionEq_3_2}. The nuclear Green's functions are then defined as follows, the trace being taken with respect to the nuclear BO states $\ket{\chi}$, namely
\begin{equation} 
D^{BO}_{\alpha_{\bar{n}}}\left(k_{\bar{n}}t_{\bar{n}}\right) \equiv \frac{1}{i^{n-1}} \frac{\text{Tr}\left[e^{-\beta \hat{H}_{n}} \mathcal{T}\left\{ \hat{R}_{\alpha_{\bar{n}}}\left(k_{\bar{n}}t_{\bar{n}}\right) \right\} \right]_{\chi}}{\text{Tr}\left[e^{-\beta \hat{H}_{n}}\right]_{\chi}},
\label{eq:BornOppenheimerTheoryEq_6}
\end{equation}
where we use the notation $\alpha_{\bar{n}} \equiv \alpha_{1} \cdots \alpha_{n}$, $k_{\bar{n}}t_{\bar{n}} \equiv k_{1}t_{1},\ldots, k_{n}t_{n}$ and $\hat{R}_{\alpha_{\bar{n}}}\left(k_{\bar{n}}t_{\bar{n}}\right) \equiv \hat{R}_{\alpha_{1}}\left(k_{1}t_{1}\right) \cdots \hat{R}_{\alpha_{n}}\left(k_{n}t_{n}\right)$. Here the operators like $\hat{R}_{\alpha}\left(kt\right) = \hat{U}^{\dagger}_{n}\left(t\right) \hat{R}_{\alpha}\left(k\right) \hat{U}_{n}\left(t\right)$ are operators in the Heisenberg picture defined with respect to the Hamiltonian $\hat{H}_{n}$. The nuclear variables satisfy the canonical commutation relation
\begin{equation}
\left[\hat{R}_{\alpha}\left(kt\right),\hat{P}_{\beta}\left(k't\right)\right]_{-} = i \delta_{\alpha\beta} \delta_{kk'}.
\label{eq:BornOppenheimerTheoryEq_7}
\end{equation}

\subsection{Non-Reduced Formulation}
\label{NonReducedFormulation}

The first many-body Green's function theory based on the non-reduced Hamiltonian of Eq. \ref{eq:IntroductionEq_1} as such was developed by Baym in the 1960's \cite{Baym-field-1961}. Here, the harmonic approximation was made for the nuclear variables. Namely, the nuclei were assumed to be rather well localized in certain regions of space and the nuclear variable dependent potential parts were expanded up to second order in displacements from the equilibrium positions of the nuclei. Given the Hamiltonian, written in second quantization for electrons and in first quantization for nuclei, the Green's functions are defined as in Eqs. \ref{eq:BornOppenheimerTheoryEq_4} and \ref{eq:BornOppenheimerTheoryEq_6}, but in this case with respect to the states $\ket{\Psi}$ in the full electron-nuclear Hilbert space and with respect to the Hamiltonian $\hat{H}$. The equations of motion (EOM) for these quantities are then derived and by solving the EOM, physical observables can be computed. The theory of Baym has been further considered in several studies \cite{Keating-DielScreeningAndThePhononSpectraOfMetalAndNonMetalCrystals-PhysRev.175.1171-1968,Giustino-ElectronPhononInteractFromFirstPrinc-RevModPhys.89.015003-2017,Stefanucci-InAndOutOfEquilibriumAbInitioTheoryOfElectronsAndPhonons-PhysRevX.13.031026-2023} and extended also beyond the harmonic approximation \cite{Gillis-SelfConsistentPhononsAndTheCoupledElectronPhononSystem-PhysRevB.1.1872-1970}. All these approaches are based on the non-reduced Hamiltonian of Eq. \ref{eq:IntroductionEq_1} and thus molecules with rotations and rotational-vibrational coupling are not accessible. We also have to be careful when computing observables as unphysical results will in many cases occur, if the issues originating from the symmetry are not handled correctly. In Sec. \ref{SymmetryAndReduction}, we considered the origin of these issues and gave examples how they can manifest in practice.

\subsection{Reduced Formulation}
\label{ReducedFormulation}

Here we consider the reduced formulation (also called the body-fixed formulation) of the problem defined by the non-reduced Hamiltonian of Eq. \ref{eq:IntroductionEq_1} developed in \cite{Harkonen-ManyBodyGreensFunctionTheoryOfElectronsAndNucleiBeyondTheBornOppenheimerApproximation-PhysRevB.101.235153-2020}. The first steps for the many-body Green's function theory based on the reduced Hamiltonian were taken in \cite{vanLeeuwen-FirstPrincElectronPhonon-PhysRevB.69.115110-2004}, but not all terms originating from the reduced Hamiltonian were included to the EOM and no EOM for the nuclear variables was considered. The first step is to derive the reduced Hamiltonian and it is obtained by transforming \cite{Harkonen-ManyBodyGreensFunctionTheoryOfElectronsAndNucleiBeyondTheBornOppenheimerApproximation-PhysRevB.101.235153-2020}
\begin{equation} 
\vec{r}'_{i} = \mathcal{R}\left(\boldsymbol{\theta}\right) \left( \vec{r}_{i} - \vec{R}_{cmn} \right), \quad \vec{R}'_{k} = \vec{R}_{k} - \vec{R}_{cmn}, \quad \vec{R}'_{N_{n}} = \vec{R}_{cm},
\label{eq:ReducedFormulationEq_1}
\end{equation}
where $i = 1,\ldots,N_{e}$, $k = 1,\ldots,N_{n}-1$, $\mathcal{R}\left(\boldsymbol{\theta}\right)$ is the rotation matrix, the nuclear center-of-mass coordinate is $\vec{R}_{cmn} = \frac{1}{ M_{nuc} } \sum_{ k } M_{k} \vec{R}_{k}$, the total nuclear mass is $M_{nuc} \equiv \sum_{ k } M_{k}$ and $\vec{R}_{cm}$ is the center-of-mass coordinate of the system. We note that this was a particular choice of $\left(\vec{r}', \vec{R}'\right)$ and there are many other such choices \cite{Sutcliffe-TheDecouplingOfElectronicAndNuclearMotions-2000}. The Hamiltonian in the coordinates $\left(\vec{r}', \vec{R}'\right)$ can be written as in Eq. \ref{eq:ReductionEq_1}, $H = T_{cm} + H'$ \cite{Harkonen-ManyBodyGreensFunctionTheoryOfElectronsAndNucleiBeyondTheBornOppenheimerApproximation-PhysRevB.101.235153-2020}. We can thus study independently the internal properties of the system described by the reduced Hamiltonian $H'$ and of the center-of-mass of the system which behaves as a free particle. The Green's functions are then defined as in the non-reduced formulation but here with respect to the reduced Hamiltonian $H'$ and corresponding states of the reduced space $\ket{\Psi'}$, namely
\begin{align} 
G\left(\vec{y}t,\vec{y}'t'\right) &\equiv \frac{1}{i} \frac{\text{Tr}\left[e^{-\beta \hat{H}'^{M}} \mathcal{T}\left\{ \hat{\psi}\left(\vec{y}t\right) \hat{\psi}^{\dagger}\left(\vec{y}'t'\right) \right\} \right]_{\Psi'}}{\text{Tr}\left[e^{-\beta \hat{H}'^{M}}\right]_{\Psi'}}, \nonumber \\
D_{\alpha_{\bar{n}}}\left(k_{\bar{n}}t_{\bar{n}}\right) &\equiv \frac{1}{i^{n-1}} \frac{\text{Tr}\left[e^{-\beta \hat{H}'^{M}} \mathcal{T}\left\{ \hat{R}_{\alpha_{\bar{n}}}\left(k_{\bar{n}}t_{\bar{n}}\right) \right\} \right]_{\Psi'}}{\text{Tr}\left[e^{-\beta \hat{H}'^{M}}\right]_{\Psi'}}.
\label{eq:ReducedFormulationEq_3}
\end{align}
The EOM for the Green's functions can be written as (a special case $n = 2$ for nuclear Green's function given) \cite{Harkonen-ManyBodyGreensFunctionTheoryOfElectronsAndNucleiBeyondTheBornOppenheimerApproximation-PhysRevB.101.235153-2020}
\begin{align} 
\delta\left(t-t'\right)\delta\left(\vec{y} - \vec{y}'\right) &= \left[i \frac{\partial}{\partial{t}} + \frac{ \nabla^{2}_{\vec{y}} }{2 m_{e}}  - V_{tot}\left(\vec{y} t\right) \right] G\left(\vec{y} t,\vec{y}' t'\right) - \int d\vec{y}'' \int dt''  \Sigma\left(\vec{y} t,\vec{y}'' t''\right) G\left(\vec{y}'' t'',\vec{y}' t'\right), \nonumber \\
M_{k} \frac{\partial^{2}{}}{\partial{t^{2}}} D_{\alpha\beta}\left(kt,k't'\right) &= - \sum_{k'',\alpha'} \int dt'' \Pi_{\alpha \alpha'}\left(k t,k''t''\right) D_{\alpha' \beta }\left(k''t'',k't'\right) - \delta_{\alpha\beta} \delta_{kk'} \delta\left(t-t'\right).
\label{eq:ReducedFormulationEq_4}
\end{align} 
If we can solve the Green's function from the equations of motion of Eq. \ref{eq:ReducedFormulationEq_4} or by directly evaluating from Eq. \ref{eq:ReducedFormulationEq_3} by using many-body perturbation theory, for instance, we can compute essentially the expected value of any observable. However, this is currently too complicated task and further approximations are needed. We note that even though the BO Green's functions of Sec. \ref{BornOppenheimerTheory} have been solved with numerical approaches \cite{Golze-TheGWCompendiumAPracticalGuideToTheoreticalPhotoemissionSpectroscopy-2019}, the Green's functions of Eq. \ref{eq:ReducedFormulationEq_3} are more general objects defined with respect to the full electron-nuclear states. For this reason we do not have readily available computational packages to compute these quantities in practice. Partially for this reason, we have combined the so-called exact factorization and the many-body Green's function approaches. We summarize this approach in Sec. \ref{ExactFactorizationOfGreensFunctions}.

\subsection{Exact Factorization of Green's Functions}
\label{ExactFactorizationOfGreensFunctions}

In the BO approximation we approximated the full wave function $\Psi\left(\vec{r}, \vec{R}\right)$ as a simple product. It turns out, however, that such a factorization can be shown to be exact \cite{Hunter-ConditionalProbInWaveMech-1975,Gidopoulos-ElectronicNonAdiabaticStates-2005,Gidopoulos-Gross-ElectronicNonAdiabaticStates-2014} such that
\begin{equation}
\Psi\left(\vec{r},\vec{R}\right) = \Phi_{\vec{R}}\left(\vec{r}\right) \chi\left(\vec{R}\right),
\label{eq:ExactFactorizationOfGreensFunctionsEq_1}
\end{equation}
where the electronic wave function $\Phi_{\vec{R}}\left(\vec{r}\right)$ and the nuclear wave function $\chi\left(\vec{R}\right)$ satisfy \cite{Gidopoulos-Gross-ElectronicNonAdiabaticStates-2014}
\begin{equation} 
H_{n} \chi\left(\vec{R}\right) = E \chi\left(\vec{R}\right), \quad H_{e} \Phi_{\vec{R}}\left(\vec{r}\right) = \epsilon\left(\vec{R}\right) \Phi_{\vec{R}}\left(\vec{r}\right).
\label{eq:ExactFactorizationOfGreensFunctionsEq_2}
\end{equation}
For simplicity, we have given the non-reduced equations, where the Hamiltonians are
\begin{equation} 
H_{n} = \sum_{k} \frac{ 1 }{2M_{k}} \left[-i \nabla_{\vec{R}_{k}} + \vec{A}_{k}\left(\vec{R}\right) \right]^{2} + \epsilon\left(\vec{R}\right), \quad  H_{e} = H_{BO}\left(\vec{r},\vec{R}\right) +  U_{en}\left(\vec{R}\right),
\label{eq:ExactFactorizationOfGreensFunctionsEq_3}
\end{equation}
and the scalar and vector potentials are of the form $\epsilon\left(\vec{R}\right) = \int d\vec{r} \Phi^{\ast}_{\vec{R}}\left(\vec{r}\right) H_{e} \Phi_{\vec{R}}\left(\vec{r}\right)$ and $\vec{A}_{k}\left(\vec{R}\right) = -i \int d\vec{r} \Phi^{\ast}_{\vec{R}}\left(\vec{r}\right)  \nabla_{\vec{R}_{k}} \Phi_{\vec{R}}\left(\vec{r}\right)$. In Eq. \ref{eq:ExactFactorizationOfGreensFunctionsEq_3}, the operator $U_{en}$ is acting on the nuclear variables only and is of the form
\begin{equation} 
U_{en}\left(\vec{R}\right) = \sum_{k} \frac{ 1 }{2 M_{k} } \left[ \left(-i \nabla_{\vec{R}_{k}} - \vec{A}_{k}\right)^{2} + 2 \left( \vec{D}_{k} + \vec{A}_{k}\right) \cdot \left(-i \nabla_{\vec{R}_{k}} - \vec{A}_{k}\right) \right],
\label{eq:ExactFactorizationOfGreensFunctionsEq_5}
\end{equation}
where $\vec{D}_{k}\left(\vec{R}\right) = -i \chi^{-1}\left(\vec{R}\right) \nabla_{\vec{R}_{k}} \chi\left(\vec{R}\right)$. The wave functions in exact factorization are normalized as $\int d\vec{R} \left| \chi\left(\vec{R}\right) \right|^{2} = \int d\vec{r} \left|\Phi_{\vec{R}}\left(\vec{r}\right)\right|^{2} = 1$. The solution of the exact factorized equations in Eq. \ref{eq:ExactFactorizationOfGreensFunctionsEq_2} provide an exact and alternative way to solve the original Schr\"{o}dinger equation given by Eq. \ref{eq:IntroductionEq_2}. Interestingly, a special case where the vector potential $\vec{A}_{k}$ and the potential $U_{en}$ vanish, the exact equations of Eq. \ref{eq:ExactFactorizationOfGreensFunctionsEq_2} become the BO equations of Eqs. \ref{eq:IntroductionEq_3_1} and \ref{eq:IntroductionEq_3_2}. The exact factorization thus gives a way to generate beyond-BO approximations in a systematic way.

We recently combined the exact factorization and many-body Green's function theory \cite{Harkonen-ExactFactorizationOfTheManyBodyGreensFunctionTheoryOfElectronsAndNuclei-PhysRevB.106.205137-2022} and call this an exact factorization of many-body Green's functions. This approach provides an alternative way of writing the exact Green's functions of Eq. \ref{eq:ReducedFormulationEq_3} and allows a starting point for computationally accessible approximation. We showed \cite{Harkonen-ExactFactorizationOfTheManyBodyGreensFunctionTheoryOfElectronsAndNuclei-PhysRevB.106.205137-2022} that the BO Green's functions of Eqs. \ref{eq:BornOppenheimerTheoryEq_4} and \ref{eq:BornOppenheimerTheoryEq_6} follow as the simplest special case of the factorized exact functions. The more complicated approximation takes into account the coupling of the electronic and nuclear problems in a more sophisticated ways. Overall, we expect this approach to be useful in practical implementations as in many cases the quantities involved in the expansion can be obtained from the corresponding BO theory. We can also study the beyond-BO contributions in a systematic fashion the lowest order contributions being relatively simple, when compared to the exact solution of the problem.

\subsection{Quantum Field Theory of Electrons and Nuclei}
\label{QuantumFieldTheoryOfElectronsAndNuclei}

In all Green's function approaches presented so far, the electronic variables were treated in terms of field operators while the nuclear variables in first quantization, in terms of operators $\hat{\vec{R}}$ whose representatives are $\vec{R}$. This is often well justified as the nuclei are usually rather well localized near their equilibrium positions. This means that we usually conduct a transformation $\hat{\vec{R}} = \vec{x} + \hat{\vec{u}}$, where $\vec{x}$ are constants called the reference positions. These quantities, in principle rather arbitrary, can be chosen with the criteria $\vec{x}  = \langle \hat{\vec{R}} \rangle$ after which they are called the equilibrium positions \cite{Harkonen-ManyBodyGreensFunctionTheoryOfElectronsAndNucleiBeyondTheBornOppenheimerApproximation-PhysRevB.101.235153-2020}. This transformation is often useful in the sense that we usually expand the Hamiltonian in $\hat{\vec{u}}$ about the reference positions $\vec{x}$ and thus obtain a relatively simple theory for the nuclear variables when the expansion is restricted to second order. Again, this procedure is justified by the well localized nature of the nuclei. However, there are systems, like quantum crystals \cite{Cazorla-SimulationAndUnderstandingOfAtomicAndMolecularQuantumCrystals-RevModPhys.89.035003-2017}, where this is not necessarily true. Also with systems having significant anharmonicity, the lowest order expansion in $\hat{\vec{u}}$ is not valid, which renders the theory very complex due to the higher order anharmonic terms. To describe these kind of systems, we reformulated the Coulomb problem as a quantum field theory \cite{Harkonen-QuantumFieldTheoryOfElectronsAndNuclei-2024}. For simplicity, the first formulation was done for the non-reduced Hamiltonian, but in particular for crystals with periodic boundary conditions, the center-of-mass reduced version will have similarities to the non-reduced one \cite{Harkonen-ManyBodyGreensFunctionTheoryOfElectronsAndNucleiBeyondTheBornOppenheimerApproximation-PhysRevB.101.235153-2020}. Additionally, further approximations to the exact equations of motion, some of which must eventually be made to make the theory computationally feasible in practice, may render the non-reduced formulation useful.

The non-reduced quantum field theory of electrons and nuclei can be obtained by considering the field Lagrangian
\begin{align} 
L =& \sum^{N_{s}}_{s = 1} \int d\vec{y} \psi^{\dagger}_{s}\left(\vec{y}\right) \tilde{D}_{s}\left(\vec{y}\right) \psi_{s}\left(\vec{y}\right) - \frac{1}{2} \sum^{N_{s}}_{s = 1} \int d\vec{y} \int d\vec{y}'  v_{ss}\left(\vec{y},\vec{y}'\right) \psi^{\dagger}_{s}\left(\vec{y}\right) \psi^{\dagger}_{s}\left(\vec{y}'\right) \psi_{s}\left(\vec{y}'t\right) \psi_{s}\left(\vec{y}\right) \nonumber \\
&- \frac{1}{2} \sum^{N_{s}}_{s , s' = 1}{}^{'} \int d\vec{y} \int d\vec{y}'  v_{ss'}\left(\vec{y},\vec{y}'\right) n_{s}\left(\vec{y}\right) n_{s'}\left(\vec{y}'\right),
\label{eq:QuantumFieldTheoryOfElectronsAndNucleiEq_1}
\end{align}
where $n_{s}\left(\vec{y}\right) \equiv \psi^{\dagger}_{s}\left(\vec{y}\right) \psi_{s}\left(\vec{y}\right)$, $\tilde{D}_{s}\left(\vec{y}t\right) \equiv i \frac{\partial{}}{\partial{t}} + \frac{ 1 }{2 m_{s}} \nabla^{2}_{s}$ and $v_{ss'}\left(\vec{y},\vec{y}'\right) \equiv Z_{s} Z_{s'} \chi_{ss'} v\left(\vec{y}, \vec{y}'\right)$. Here $s$ and $s'$ label the different species of particles, electrons and different types of nuclei. The Lagrangian given by Eq. \ref{eq:QuantumFieldTheoryOfElectronsAndNucleiEq_1} generalizes the non-relativistic quantum field theory given in Ref. \cite{Baiguera-AspectsOfNonRelativisticQuantumFieldTheories-2024} to include more than one species of particles. After the canonical quantization, the Schrödinger fields $\psi_{s}\left(\vec{y}\right), \psi^{\dagger}_{s}\left(\vec{y}\right)$ become operators satisfying the following (anti)commutation relations
\begin{equation} 
\left[\hat{\psi}_{s}\left(\vec{y}t\right),\hat{\psi}^{\dagger}_{s}\left(\vec{y}'t\right)\right]_{\pm} =  \delta\left(\vec{y}-\vec{y}'\right),
\label{eq:QuantumFieldTheoryOfElectronsAndNucleiEq_3}
\end{equation}
with all the remaining inter/intra-species anti-commutators and/or commutators vanishing \cite{Harkonen-QuantumFieldTheoryOfElectronsAndNuclei-2024}. We define the one-body Green's function for a species $s$ as
\begin{equation} 
G_{s}\left(\vec{y}t,\vec{y}'t'\right) \equiv \frac{1}{i} \frac{\text{Tr}\left[e^{-\beta \hat{H}^{M}} \mathcal{T}\left\{ \hat{\psi}_{s}\left(\vec{y}t\right) \hat{\psi}^{\dagger}_{s}\left(\vec{y}'t'\right) \right\} \right]_{\Psi}}{\text{Tr}\left[e^{-\beta \hat{H}^{M}}\right]_{\Psi}},
\label{eq:QuantumFieldTheoryOfElectronsAndNucleiEq_4}
\end{equation}
where the operators are in the Heisenberg picture. The equations of motion can be written in the form of the Hedin's equations \cite{Schwinger-OnTheGreensFunctionsOfQuantizedFieldsII-1951,Hedin-NewMethForCalcTheOneParticleGreensFunctWithApplToTheElectronGasProb-PhysRev.139.A796-1965}, namely \cite{Harkonen-QuantumFieldTheoryOfElectronsAndNuclei-2024}
\begin{eqnarray} 
\delta\left(1 - 2\right) &=& \left[i \hbar\frac{\partial}{\partial{t_{1}}} + \frac{\hbar^{2} }{2 m_{s}} \nabla^{2}_{s} - V_{tot}\left(1,s\right) \right] G_{s}\left(1,2\right) - \int d3  \Sigma_{s}\left(1,3\right) G_{s}\left(3,2\right), \nonumber \\
\Sigma_{s}\left(1,4\right)  &=& i \hbar \int d3 \int d5 W_{s}\left(1,5\right)  G_{s}\left(1,3\right)  \Gamma_{s}\left(3,4,5\right), \nonumber \\
\Gamma_{s}\left(1,2,3\right)&=& \delta\left(1 - 2\right) \delta\left(1 - 3\right) + \int d4\int d5 \int d6 \int d7 \frac{\delta{\Sigma_{s}\left(1,2\right)}}{\delta{G_{s}\left(4,5\right)}} G_{s}\left(4,6\right) G_{s}\left(7,5\right) \Gamma_{s}\left(6,7,3\right), \nonumber \\
 			W_{s}\left(1,2\right) &=& Z^{2}_{s} v\left(1,2\right) +  Z^{2}_{s} \int d3 \int d4  v\left(2,3\right) \sum_{s'} P_{s'}\left(3,4\right)  W_{s'}\left(1,4\right), \nonumber \\
P_{s}\left(1,2\right)       &=& -i \hbar \int d3 \int d4  G_{s}\left(1,3\right) G_{s}\left(4,1^{+}\right) \Gamma_{s}\left(3,4,2\right).
\label{eq:QuantumFieldTheoryOfElectronsAndNucleiEq_5}
\end{eqnarray}
Here we use the following shorthand notations: $1 = \vec{y} t, \ \delta\left(1-2\right) = \delta\left(t-t'\right)\delta\left(\vec{y} - \vec{y}'\right)$ and so on. The coupling of different species is through all terms of Eq. \ref{eq:QuantumFieldTheoryOfElectronsAndNucleiEq_5}: the screened Coulomb interaction $W_{s}\left(1,2\right)$, the self-energy $\Sigma_{s}\left(1,4\right)$, the vertex function $\Gamma_{s}\left(1,2,3\right)$ and the polarization $P_{s}\left(1,2\right)$. The equations of motion written as in Eq. \ref{eq:QuantumFieldTheoryOfElectronsAndNucleiEq_5} can be beneficial. Namely, there is a lot of technical know how to solve such a set of equations within the electronic BO approximation \cite{Golze-TheGWCompendiumAPracticalGuideToTheoreticalPhotoemissionSpectroscopy-2019} and the equations for all species of particles formally assume the same form. Moreover, a formulation as a quantum field theory makes it easier to apply the sophisticated theoretical machinery used in non-relativistic quantum field theories \cite{Baiguera-AspectsOfNonRelativisticQuantumFieldTheories-2024} to the electron-nuclei problem.

\section{Conclusions}
\label{Conclusions}

We have seen that the reduction theory, relevant aspect in the solution of gravitational $n$-body problem, is of central importance in the solution of quantum mechanical many-body problem of Coulomb interactions. These considerations have been essentially neglected in the solid state physics community the reasons being likely related to the BO approximation, essentially always imposed, and at the same time removing some of the problems due to the symmetry. In the case of molecules, the wave function approach developed by Born and Oppenheimer was already based on the reduced Hamiltonian. The dimensionality problem prevents us to use the wave function approach in solids and this creates a need for alternative approaches like Green's function theory. The non-reduced nature of the earliest beyond-BO Green's function theories has motivated the development of the reduced theory, valid for molecules and solids, both within and beyond the BO approximation.

In this work, aside from summarizing the different Green's function theories of the Coulomb problem, we make an effort to bridge the seemingly separate disciplines of physics, which are confronted with the same issues. The debate on relational versus absolute description of nature has lasted for over two thousand years, starting with Aristotle, continued by Descartes, Newton and Leibnitz, and noted by Mach, whose principle played a part in the development of general relativity by Einstein \cite{Einstein-DieGrundlagenDerAllgemeinen-1916,Barbour-TheDiscoveryOfDynamicsAStudyFromAMachianPointOfViewOfTheDiscoveryAndTheStructureOfDynamicalTheories-2001}. The debate continues and is surfacing in one way or the other, sometimes quite implicitly, in many aspects of modern physics \cite{Rovelli-RelationalQuantumMechanics-1996,Barbour-RelationalConceptsOfSpaceAndTime-1982,Dickson-AviewFromNowhereQuantumReferenceFramesAndUncertainty-2004,Butterfield-OnSymplecticReductionInClassicalMechanics-2006,Elze-LinearDynamicsOfQuantumClassicalHybrids-PhysRevA.85.052109-2012,Diosi-CentreOfMassDecoherenceDueToTimeDilationParadoxicalFrameDependence-2017,Smolin-TemporalRelationalism-2018,Ahmad-QuantumRelativityOfSubsystems-PhysRevLett.128.170401-2022,Harkonen-BreakdownOfTheBornOppenheimerApproximationInSolidHydrogenAndHydrogenRichSolids-Arxiv-2023}. We find this topic central to many-body Green's function theories, a subject that, to the best of our knowledge, has not been discussed within the many-body solid-state physics community. Based on our discussion, we conclude that both, absolute and relational facets of our models of nature are playing an important role. Namely, the absolute view is necessary and follows from the requirement of global conservation laws and thus global symmetries of isolated systems. The relational view is important in formulating the theory in such a form that meaningful observables can be computed. We see both of these aspects as essential in any exact quantum mechanical description of isolated systems, where the system itself is considered to be the whole universe.

\bibliography{bibfile}
\bibliographystyle{phpf}

\end{document}